\documentclass[lettersize,journal]{IEEEtran}

\usepackage{bm}
\usepackage{enumerate}
\usepackage{enumitem}
\usepackage{makecell}
\usepackage{multirow}
\usepackage{stfloats}
\usepackage{amsmath}
\usepackage{color}
\usepackage{amsfonts}
\usepackage{mathtools} 
\usepackage{amssymb}
\usepackage{verbatim}
\usepackage[normalem]{ulem}
\useunder{\uline}{\ul}{}
\usepackage{booktabs}
\usepackage{dsfont}
\usepackage[table]{xcolor}

\usepackage{algorithm}
\usepackage{algorithmic}
\usepackage{enumerate}

\allowdisplaybreaks[3]

\begin{document}

\title{Modality Reliability Guided Multimodal Recommendation}

\author{Xue~Dong, Xuemeng Song~\IEEEmembership{Senior Member,~IEEE}, Na Zheng, Sicheng Zhao~\IEEEmembership{Senior Member,~IEEE}, Guiguang Ding
\IEEEcompsocitemizethanks{
\IEEEcompsocthanksitem X. Dong and G. Ding are with the School of Software, Tsinghua University, Beijing 100084, China. E-mail: dongxue.sdu@gmail.com, dinggg@tsinghua.edu.cn.
\IEEEcompsocthanksitem X. Song is with the School of Computer Science and Technology, Shandong University, Qingdao 266237, China. E-mail: sxmustc@gmail.com.
\IEEEcompsocthanksitem N. Zheng is with the with Grab-NUS AI Lab, National University of Singapore, Singapore 117602. E-mail: zhengnagrape@gmail.com. 
\IEEEcompsocthanksitem S. Zhao is with the Beijing National Research Center for Information Science and Technology (BNRist), Tsinghua University, Beijing 100084, China. E-mail: schzhao@gmail.com. 
}
}

\markboth{Journal of \LaTeX\ Class Files,~Vol.~14, No.~8, August~2021}%
{Shell \MakeLowercase{\textit{et al.}}: A Sample Article Using IEEEtran.cls for IEEE Journals}

\IEEEtitleabstractindextext{%
\begin{abstract}
Multimodal recommendation faces an issue of the performance degradation that the uni-modal recommendation sometimes achieves the better performance. A possible reason is that the unreliable item modality data hurts the fusion result. Several existing studies have introduced weights for different modalities to reduce the contribution of the unreliable modality data in predicting the final user rating. However, they fail to provide appropriate supervisions for learning the modality weights, making the learned weights imprecise. Therefore, we propose a modality reliability guided multimodal recommendation framework that uniquely learns the modality weights supervised by the modality reliability. 
Considering that there is no explicit label provided for modality reliability, we resort to automatically identify it through the BPR recommendation objective. In particular, we define a modality reliability vector as the supervision label by the difference between modality-specific user ratings to positive and negative items, where a larger difference indicates a higher reliability of the modality as the BPR objective is better satisfied. Furthermore, to enhance the effectiveness of the supervision, we calculate the confidence level for the modality reliability vector, which dynamically adjusts the supervision strength and eliminates the harmful supervision. Extensive experiments on three real-world datasets show the effectiveness of the proposed method.
\end{abstract}

}

\maketitle

\IEEEdisplaynontitleabstractindextext

\IEEEpeerreviewmaketitle

\section{Introduction}
\IEEEPARstart{T}{raditional} recommendation systems~\cite{lightgcn,layergcn} model the user interests based on the user-item behaviors on the web. 
Beyond the user behaviors, the item multimodal data, e.g., images and texts, provides a comprehensive manner for capturing the user interests. 
Accordingly, many researches have thus resorted to the multimodal recommendation systems~\cite{mmgcn, grcn, JiLZWNW23, WeiLLWNC23, duple, GuoL0WSR24} that jointly exploit user behaviors and the item multimodal data to capture the user interests.

\begin{figure}[!t]
\centering
\setlength{\abovecaptionskip}{0.1cm}
\setlength{\belowcaptionskip}{-0.5cm}
\includegraphics[width=\linewidth]{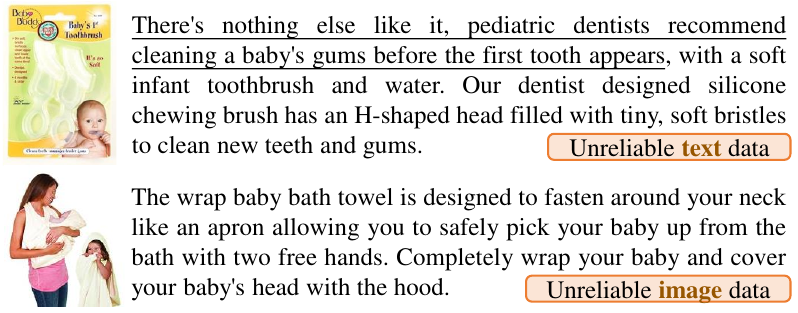}
\caption{Examples of items with unreliable modality data.}
\label{fig_motivation}
\end{figure}

Although the multimodal data can be helpful, several studies~\cite{mgcn, dragon, ZhangLLWW24} have found an crucial issue of the model performance degradation: the uni-modal recommendation framework sometimes achieves better performance than the multimodal one.
One possible reason may be that certain modality data of an item is unreliable, which hurts the multimodal fusion result. For example, as shown in Figure~\ref{fig_motivation}, the first item is a ``silicone chewing brush'', while its description contains some irrelevant information (see underlined sentenses). Besides, in the image of the second item, the mother that hugs a baby and a little girl that stands by are irrelevant to this ``wrap baby bath towel''. 
Thus, the irrelevant information in the modality data could hinder the feature extraction and negatively affect the multimodal fusion.

Several research efforts~\cite{dualgnn, dragon} have introduced weights for modalities of the item to reduce the contribution of the unreliable modality data in predicting the final user rating.  
They first calculate the modality-specific user ratings to one item in all modalities, and then fuse them based on the weights to predict the final user rating. The weights are randomly initialized and optimized with only the recommendation objective, e.g., Bayesian Personal Ranking (BPR) loss~\cite{bpr}. 
Despite their effectiveness, the modality weights learned by existing methods may be imprecise due to the lack of explicit supervisions for the weight learning. In other words, the modality weights may be not optimized as expected, but as additional parameters to increase the performance.

\begin{figure}[!t]
\centering
\setlength{\abovecaptionskip}{0.1cm}
\setlength{\belowcaptionskip}{-0.5cm}
\includegraphics[width=\linewidth]{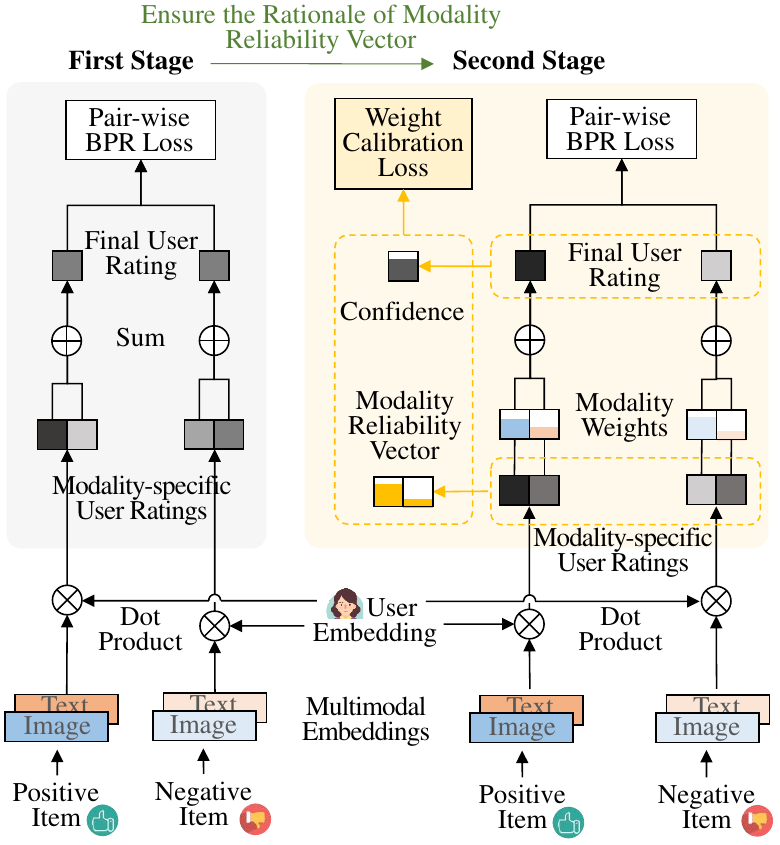}
\caption{Illustration of the proposed MARGO framework, which adds explicit supervisions to the modality weight learning. Specifically, we introduce a new modality reliability vector through the training objective to simulate the reliability of the modality data. We also involve the confidence for the modality reliability vector to ensure the effectiveness of the supervision.}
\label{fig_model}
\end{figure}

Therefore, in this paper, we present a novel \underline{m}od\underline{a}lity \underline{r}eliability \underline{g}uided multimodal rec\underline{o}mmendation framework, termed as MARGO, shown in Figure~\ref{fig_model}. 
MARGO follows the standard multimodal recommendation framework that the modality-specific user ratings are fused with modality weights and the pair-wise BPR loss is adopted for model optimization. 
Differently, MARGO introduces the supervision of the modality reliability on the modality weight learning for a better fusion result. 
To infer the modality reliability, we draw inspiration from the BPR loss that the user ratings to positive items should be higher than negative items. Therefore, if a modality-specific user rating to the positive item is lager than that to a negative one, the corresponding modality of the two items tend to be reliable for predicting the final user rating. On the contrary, the modality of at least one in the two items is unreliable as it conflicts the BPR objective. 
Consequently, we define a modality reliability vector by the difference between the modality-specific user ratings to the positive and negative items, which can be used as the supervision of the modality weights of the two items.
However, directly adding such supervision may be flawed, since the modality reliability vector is not always confident. 
To tackle this, inspired by that the loss value usually reflects the confidence of the result, we calculate the confidence for the modality reliability vector based on the BPR loss value, i.e., the difference between the final user ratings to the positive and negative items, where a smaller loss indicates a higher confidence. 
Thereafter, we design a weight calibration loss that pushes up the similarity between the modality reliability vector and the modality weights, where the supervision strength is dynamically adjusted by the confidence.

Besides, we design a two-stage training process to ensure the rationale of the modality reliability vector and the effectiveness of the supervision. The first stage discards the modality weights and pre-trains the model with only the BPR loss. 
After the model converge, the predicted modality-specific user ratings are meaningful and thus the modality reliability vector is rational. 
Then the second stage involves the modality weights and fine-tunes the model by incorporating the BPR loss and the weight calibration loss. We conduct extensive experiments on three public datasets, and the results have proven that MARGO obtains superior performances compared to state-of-the-art methods.
The main contributions can be summarized as follows.
\begin{itemize}[leftmargin=3mm]
\item We propose a modality reliability guided multimodal recommendation framework that learns the modality weights with the supervision of the modality reliability vector. It is the first attempt to add explicit supervisions to the weight learning during the multimodal fusion. 
\item We propose an automatic method to derive the supervision label for the modality weights, where we define a modality reliability vector through the BPR loss.
\item We design a weight calibration loss, which dynamically adjusts the supervision strength of the modality reliability vector on the weights and eliminates the harmful supervision through a newly introduced confidence.
\item Extensive experiments on the three public datasets have demonstrated the effectiveness of the proposed method. We will release our codes to facilitate other researchers in https://github.com/hello-dx/MARGO.
\end{itemize}

\section{Related work}
Traditional recommendation methods~\cite{bpr,lightgcn,layergcn} utilize the historical user-item interactions to learn user interests. Considering that the multimodal information reflects the item properties from different perspectives, some researches resort to the multimodal recommendation~\cite{amcf,vbpr,mmgcn,dualgnn,lattice,ChenSGZ23, LinTZLW0WY23}. Several researches implicitly use the multimodal data as supervision signals, while other researches explicitly leverage it as the side information of items participating in the model training. This section will detail the implicit and explicit multimodal recommendation methods.

\subsection{Implicit Multimodal Recommendation}
These researches leverage the item multimodal information as the prior knowledge or supervisions to improve the recommendation performance~\cite{amcf, lattice, freedom, ZhouZLZMWYJ23}.
For example, AMCF~\cite{amcf} learns a projection between the item ID embedding and item attribute embeddings. The item attributes are adopted to distillate the semantic information into the item ID embedding, without directly participating in the user rating prediction. Zhang~et~al.~\cite{lattice} and Zhou~et~al.~\cite{freedom} adopt the item multimodal features to calculate the prior similarities between items. Based on the similarity, they construct a homogeneous graph between items and enhance the item embedding learning through the graph convolution in the graph. BM3~\cite{ZhouZLZMWYJ23} introduces inter- and intra-modality contrastive learning to align item features in different modalities. 

\subsection{Explicit Multimodal Recommendation}
These researches~\cite{vbpr, mmgcn, GuoZZSC21, KimLSK22, LiuTSYH22, Li0ZZHM023} explicitly extract the modality features from the item multimodal data and predict the user rating through the modality fusion result. According to the timing of the modality fusion, existing explicit multimodal recommendation methods could be roughly divided into the early and late fusion-based methods. 

The \textbf{early fusion-based methods}~\cite{vbpr, LiuCLH19, mgcn, ZhangXMLYL24} first fuse the modality features of one item into a single embedding and then based on it predict the final user rating to the item. For example, VBPR~\cite{vbpr} is the first attempt to leverage the item image into the recommendation, which concatenates the original item ID embedding and item visual features as the new item embedding. 
MGCN~\cite{mgcn} introduces a behavior-aware feature fusion method to filter out the redundant information between
modality features
Nevertheless, this strategy cannot model the user's specific interests on different modalities~\cite{mmgcn}. 

The \textbf{late fusion-based methods}~\cite{mmgcn,dualgnn,DongWSDN20,grcn,LiuXGSQH23,SongWSFZN23} learn the user rating towards each specific modality, and then predict the user final rating to the item by fusing all the modality-specific user ratings. For example, MMGCN~\cite{mmgcn} constructs the user-item graph in each modality and predicts the user ratings to the item in different modalities. It then utilizes the mean pooling of all modality-specific user ratings as the final user rating. 
Considering the different contributions of modalities during the modality fusion, several studies~\cite{duple,LiuCCLNK23} introduce the pre-defined a weight for each item modality. In particular, Dong~et~al.~\cite{duple} set the weights for visual and textual modalities as hyper-parameters and tune the weights through validation set. 
Other studies~\cite{dualgnn, dragon} learn the modality weights by a more flexible manner, which randomly assigns the weight for each modality in different items and updates the weight through the backward propagation. 

Despite of their effectiveness, existing methods fail to add certain supervision signals when learning the modality weights. 
This may result in the learned modality weights failing to capture the
modality reliability and becoming additional parameters used to improve the performance. Therefore, in this paper, we draw inspiration from the recommendation loss and derive a modality reliability vector from it. We then devise a weight calibration loss to supervise the learning of the modality weights.

\section{Methodology}
In this section, we first provide the preliminaries of the multimodal recommendation in Subsection~\ref{section_mmrec_preliminaries}. We then detail the proposed MARGO framework and its training algorithm in Subsection~\ref{section_MARGO} and \ref{section_training}, respectively. 
Subsection~\ref{section_analysis} provides the theoretical analysis of the model convergency.

\subsection{Preliminaries}
\label{section_mmrec_preliminaries}
We detail the preliminaries of the multimodal recommendations in this subsection, including the problem formulation, model architecture, and learning objective.

\textbf{Problem Formulation.} 
Suppose that there is a set of users~$\mathcal{U}$, a set of items~$\mathcal{I}$, and a set of item modalities~$\mathcal{M}$. Each user~$u \in \mathcal{U}$ is associated with a set of interacted items~$\mathcal{I}_u$. Each item~$i \in \mathcal{I}$ has its multimodal features~$\{\boldsymbol{f}_i^m| m=1,...,|\mathcal{M}|\}$, where $\boldsymbol{f}_i^m$ denotes the feature of the $m$-th modality. 
Multimodal recommendation aims to predict the user rating to an item according to the user interactions and item multimodal features. Ultimately, the candidate items can be ranked according to the predicted ratings and top items are recommended to the user. The main notations used in this work are detailed in Table~\ref{table_notations}.

\begin{table}[t]
\centering
\renewcommand{\arraystretch}{1.3}
\caption{Summary of the Main Notations.} 
\label{table_notations}
\setlength{\tabcolsep}{1.2mm}{
\begin{tabular}{l||l}	
    \hline
    Notation & Explanation \\
    \hline\hline
    $\mathcal{U}$, $\mathcal{I}$, $\mathcal{M}$ & Sets of users, items and modalities, respectively. \\\hline
    $\bm{e}_u \in \mathbb{R}^d$ & Embedding of the user $u$.\\\hline
    $\bm{e}_{i} \in \mathbb{R}^d$ & Embedding of the item~$i$.\\\hline
    $\boldsymbol{f}_i^m$ & The $m$-th modality feature of the item $i$. \\\hline
    $\boldsymbol{w}_i \in \mathbb{R}^{|\mathcal{M}|}$ & Weight vector of the item $i$. \\\hline
    $y_{ui}^m$ & The $m$-th modality-specific rating of user $u$ to item $i$. \\\hline
    $y_{ui}$ & Final rating of the user $u$ to the item $i$. \\\hline
    \multirow{2}{*}{$(u,i,k)$}  & Training triplet of a user $u$, a positive item $i$, and a  \\
     & negative item $k$. \\\hline
    $\bm{z}_{uik} \in \mathbb{R}^{|\mathcal{M}|}$ & Modality reliability vector of training triplet $(u,i,k)$. \\\hline
    $\gamma_{uik}$ & Confidence of the modality reliability vector $\bm{z}_{uik}$. \\\hline
    $\alpha, \beta$ & Trade-off hyper-parameters. \\\hline
\end{tabular}}	
\end{table}

\textbf{Model Architecture.} 
In this paper, we focus on the late fusion based multimodal recommendation paradigm, owing to the fact that it allows straightforward and flexible weight adjustment for different modalities during the user rating prediction. 
Formally, as for a given user~$u$ and an item~$i$, a late fusion based multimodal recommendation model~$\mathcal{F}$ first learns an embedding of each modality for the user and item as follows,
\begin{equation}
\label{eqn_embedding}
\boldsymbol{e}_u^m, \boldsymbol{e}_i^m \leftarrow \mathcal{F}(u, i, \boldsymbol{f}_i^m|\boldsymbol{\Theta}), m=1,...,M,
\end{equation}
where $\boldsymbol{e}_u^m \in \mathbb{R}^d$ and $\boldsymbol{e}_i^m \in \mathbb{R}^d$ are the user and item embeddings of the $m$-th modality, which capture the user interests and item properties from the $m$-th modality, respectively. $d$ is the embedding size. $\boldsymbol{\Theta}$ is the model parameters.

Consequently, the modality-specific user rating $y_{ui}^m$ can be predicted by the dot product of the learned user and item embeddings in the corresponding modality as follows,
\begin{equation}
y_{ui}^m = \boldsymbol{e}_u^m \cdot \boldsymbol{e}_i^m.
\end{equation}

Considering the different contributions of modalities, a randomly initialized weight vector~$\boldsymbol{w}_i$ is introduced for the item~$i$ to weighted fuse all the modality-specific user rating. Formally, the final user rating can be derived as follows,
\begin{equation}
\label{eqn_rating}
y_{ui} = \boldsymbol{w}_i^{\mathsf{T}} \cdot [y_{ui}^1, y_{ui}^2, ..., y_{ui}^{|\mathcal{M}|}] , 
\end{equation}
where $y_{ui}$ refers to the final rating of the user~$u$ to the item~$i$, indicating how likely the user would be interested in the item. 
Note that the weight vector~$\boldsymbol{w}_i$ is outputted by a softmax function to ensure the summation of the weights in one item to be equal to $1$.


\textbf{Learning Objective.} 
To optimize model parameters, the widely-used Bayesian Personalized Ranking (BPR) mechanism~\cite{bpr} is adopted to enforce the final user rating to a positive item to be higher than that to a negative item. 
Specifically, let $\mathcal{D} = \{(u,i,k)_n|n=1,2,...,N\}$ denote the training set of $N$ training triplets. Each training triplet~$(u,i,k)$ consists of a user~$u \in \mathcal{U}$, a positive item~$i \in \mathcal{I}_u$ that the user has historically interacted, and a negative item~$k \in \mathcal{I} \setminus \mathcal{I}_u$ that the user has not interacted, which indicates that the user~$u$ prefers item~$i$ compared with the item~$k$. 
Then the recommendation loss can be defined as follows,
\begin{equation}
\label{eqn_L_rec}
\mathcal{L}_{\rm rec} = -\sum_{(u,i,k) \in \mathcal{D}}{\rm log}(\sigma(y_{ui} - y_{uk})),
\end{equation}
where $y_{ui}$ and $y_{uk}$ are the user ratings to the positive item~$i$ and negative item~$k$, respectively. $\sigma$ is the sigmoid function. 

\subsection{The Proposed MARGO Framework}
\label{section_MARGO}
\begin{figure}[!t]
\centering
\setlength{\abovecaptionskip}{0.1cm}
\setlength{\belowcaptionskip}{-0.5cm}
\includegraphics[width=\linewidth]{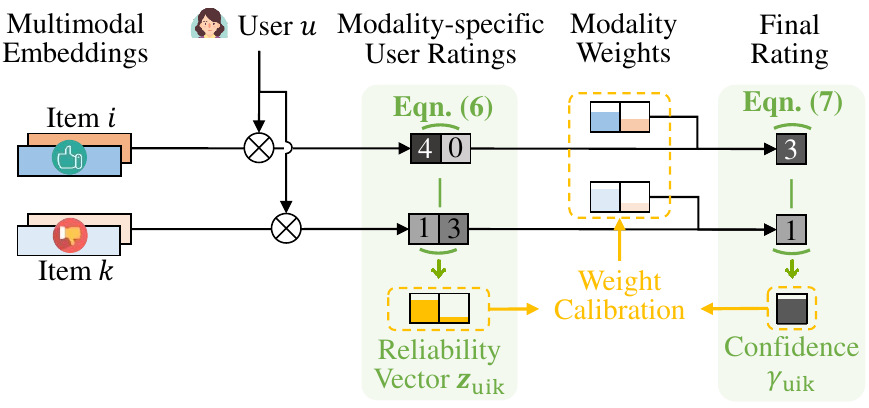}
\caption{Illustration of the weight calibration process.}
\label{fig_guiding}
\end{figure}

As a key novelty, we propose to leverage the modality reliability as the supervision signals to learn the modality weights, which makes the learned modality weights more precise and hence achieves a better multimodal fusion result. In particular, we first define a modality reliability vector through the difference between the predicted user ratings, and then introduce a weight calibration loss to supervise the weight learning process. We provide a detailed illustration of this weight calibration process in Figure~\ref{fig_guiding}.




\textbf{Modality Reliability Inference.} 
Suppose that we have a trained multimodal late fusion based recommendation model that can predict the modality-specific user ratings for all modalities in an item. Intuitively, according to BPR loss, if a modality-specific user rating to the positive item is larger than that to the negative one, we argue that the predictions in this modality are reliable and thus the modality of the two items tend to be reliable. On the contrary, this modality of at least one in the two items is unreliable as the predictions in the modality conflict with the BPR objective. 
Consequently, to a certain extent, the difference between the modality-specific user ratings to the positive and negative items could hid the information of the modality reliability, i.e., a larger difference indicates a higher reliability.

Formally, for a given training triplet $(u,i,k)$, we calculate the following difference vector through the modality-specific user ratings to the positive and negative items,
\begin{equation}
\label{eqn_difference_vector}
\boldsymbol{d}_{uik} = [y_{ui}^1 - y_{uk}^1, y_{ui}^2 - y_{uk}^2, ..., y_{ui}^M - y_{uk}^M],
\end{equation}
where $y_{ui}^m$ ($y_{uk}^m$) is the modality-specific user rating
to the $m$-th modality of the item~$i$ ($k$). 
Based on the difference vector, we define the modality reliability vector~$\boldsymbol{z}_{uik}$ of the training triplet~$(u,i,k)$ as follows,
\begin{equation}
\label{eqn_reliable_modality}
\boldsymbol{z}_{uik} = {\rm softmax}_{m}(g(\boldsymbol{d}_{uik})),
\end{equation}
where $g()$ is a mapping function that maps the negative values to infinitesimal, i.e., $g(x)=x,{\rm if} \, x \geq 0; g(x)=-e^{6},{\rm if} \, x<0$. We add this mapping function to enforce the reliability of unreliable modality to be ~$0$. Note that $\boldsymbol{d}_{uik}$ is affected by both items~$i$ and $k$, and thus the modality reliability vector~$\boldsymbol{z}_{uik}$ captures the modality reliability of both the two items.

\textbf{Weight Calibration.} Once the modality reliability vector is derived, it can be served as the supervision of the modality weight learning. 
However, it is non-trivial because that the modality reliability vector is automatically inferred through the training objective and is not always confident, where flawed cases will negatively affect the weight learning. 
Therefore, we further learn the confidence for the modality reliability vector to dynamically adjust its supervision strength on the weight learning and eliminate the negative cases.

In particular, the modality reliability vector $\boldsymbol{z}_{uik}$ is ultimately affected by the user ratings $y_{ui}$ and $y_{uk}$, and therefore its confidence can depend on the confidence of the final user rating. 
Inspired by studies~\cite{XiangDH20, ShangGCJML23, DongSTH24} that identify the confidence of the predicted result through the corresponding loss value, we define the confidence of the modality reliability vector through the the confidence of the final user rating indicated by the BPR loss value. 
Specifically, in the training triplet~$(u,i,k)$, the BPR loss calculate the difference between user ratings to the positive item~$i$ and negative item~$k$. A large difference $y_{ui}-y_{uk}$ indicates that $y_{ui}$ is larger while $y_{uk}$ is smaller. Therefore, the two predicted ratings $y_{ui}$ and $y_{uk}$ are more confident since they are in line with the objective of the BPR loss, and then the derived modality reliability vector will have a high confidence. Formally, we define the confidence~$\gamma_{uik}$ of the modality reliability vector~$\boldsymbol{z}_{uik}$ as follows,
\begin{equation}
\label{eqn_confidence}
\gamma_{uik} =
\begin{cases}
{\rm tanh}((y_{ui}-y_{uk})/\tau),& y_{ui}>y_{uk}, \\
0  ,& y_{ui}\leq y_{uk},
\end{cases}
\end{equation}
where ${\rm tanh}()$ is to map the positive confidence into $(0,1)$. $\tau$ is a linear scaling parameter to adjust the sensitivity. Note that when $y_{ui}\leq y_{uk}$, the predicted final user ratings are regarded as incorrect and thus the confidence is set to $0$.

Accordingly, we combine the modality reliability vector and its confidence as the supervision signals for learning the modality weights. 
In particular, as aforementioned, $\boldsymbol{z}_{uik}$ captures the modality reliability of both items~$i$ and $k$, and thus it should supervise the modality weights of both the two items. 
Formally, we design the weight calibration loss as follows,
\begin{equation}
\label{eqn_KL_loss_refine}
\mathcal{L}_{\rm cal} = \sum_{(u,i,k) \in \mathcal{D}} {\rm nograd}(\gamma_{uik}) {\rm KL}({\rm nograd}(\boldsymbol{z}_{uik})||\boldsymbol{w}_i \oplus \boldsymbol{w}_k),
\end{equation}
where $\oplus$ refers to the element-wise summation of two vectors. Here we simply use the element-wise summation of the weight vectors $\boldsymbol{w}_i$ and $\boldsymbol{w}_k$ to indicate the joint weight vector of the item~$i$ and item~$k$. $KL()$ is the Kullback-Leibler divergence. The subscript ${\rm nograd}()$ is a stop-gradient operator that prevents the supervision signals (i.e., the modality reliability vector and its confidence) from being updated. 
In this way, the gradients of this loss will only affect the modality weight learning as expected. 

\begin{algorithm}[!t]
\caption{Training Algorithm of MARGO.}
\label{algorithm_training}
\begin{algorithmic}[1]
\REQUIRE 
The set of users~$\mathcal{U}$, set of items~$\mathcal{I}$, and the item multimodal features. The training set~$\mathcal{D}$. 
\ENSURE Model parameters~$\boldsymbol{\Theta}$ and weight matrix~$\boldsymbol{W}$.\\
\WHILE {not converged}
\STATE $\boldsymbol{e}_u^m, \boldsymbol{e}_i^m \leftarrow \mathcal{F}(u, i, \boldsymbol{f}_i^m|\boldsymbol{\Theta})$.\\
\STATE $y_{ui}^m = \boldsymbol{e}_u^m \cdot \boldsymbol{e}_i^m$.
\STATE $y_{ui} = \sum_{m=1}^M y_{ui}^m$. \\
\STATE Update model parameters~$\boldsymbol{\Theta}$ with $\mathcal{L}_{\rm I}$ in Eqn.~(\ref{eqn_final_loss_1}). \\
\ENDWHILE
\WHILE {not converged}
\STATE $\boldsymbol{e}_u^m, \boldsymbol{e}_i^m \leftarrow \mathcal{F}(u, i, \boldsymbol{f}_i^m|\boldsymbol{\Theta})$.\\
\STATE $y_{ui}^m = \boldsymbol{e}_u^m \cdot \boldsymbol{e}_i^m$.
\STATE $\boldsymbol{z}_{uik} = {\rm softmax}_{m}(g(\boldsymbol{d}_{uik}))$.\\
\STATE $y_{ui} = \sum_{m=1}^M w^m_i y_{ui}^m$.\\
\STATE Fine-tune the model parameters~$\boldsymbol{\Theta}$ and update the weight matrix~$\boldsymbol{W}$ with $\mathcal{L}_{\rm II}$ defined in Eqn.~(\ref{eqn_final_loss_2}).\\
\ENDWHILE
\end{algorithmic}
\end{algorithm}

\subsection{Training Algorithm}
\label{section_training}
Considering that at the previous training epochs, the model is still optimized and cannot provide reliable predictions yet. The derived modality reliability vector may be not meaningful for guiding the modality weight learning. Therefore, following the study~\cite{timirec}, we adopt a two-stage learning strategy to facilitate the training process, where the first stage pre-trains the model to ensure the rational of the modality reliability vector for the modality weight learning in the second stage. 

At the first stage, in order to derive the modality-specific user ratings without the effects of modality weights, we discard the weights in Eqn.~(\ref{eqn_rating}) and simply predict the final user rating with the summations of all modality-specific user ratings, i.e., $y_{ui} = \sum_{m=1}^M y_{ui}^m$. The first stage is trained with only the recommendation loss~$\mathcal{L}_{\rm rec}$ defined in Eqn.~(\ref{eqn_L_rec}) and its loss function is defined as follows,
\begin{equation}
\label{eqn_final_loss_1}
\mathcal{L}_{\rm I}=\min_{\boldsymbol{\Theta}} (\mathcal{L}_{\rm rec} + \beta||\boldsymbol{\Theta}||_2),
\end{equation}
where last term $||\boldsymbol{\Theta}||_2$ is the regularization of model parameters to avoid over-fitting. $\beta$ is to balance the two terms. 
After the model converges in the first stage, the model is able to predict the confident user ratings, and therefore we can derive the rational modality reliability vector for supervising the modality weight learning. 

At the second stage, we calculate the final user rating by the weighted summation of all modality-specific user ratings defined in Eqn.~(\ref{eqn_rating}).
The model parameters and modality weights will be fine-tuned by jointly optimizing the recommendation loss and the weight calibration loss. Formally, the loss function of the second stage is defined as follows, 
\begin{equation}
\label{eqn_final_loss_2}
\mathcal{L}_{\rm II}=\min_{\boldsymbol{\Theta},\boldsymbol{W}} (\mathcal{L}_{\rm rec} + \alpha\mathcal{L}_{\rm cal}+\beta||\boldsymbol{\Theta}||_2),
\end{equation}
where $\alpha$ and $\beta$ are to balance the three terms. $\boldsymbol{W}$ is the weight matrix that contain the weight vectors of all items. 
The detailed training algorithm is shown in Algorithm~\ref{algorithm_training}.

\subsection{Convergency Analysis}
\label{section_analysis}
It is necessary to confirm the model convergency, since that we introduce a new weight calibration loss on the learning of the weight matrix~$\boldsymbol{W}$. To fulfill this, instead of justifying the convergency of the overall loss~$\mathcal{L}_{\rm II}$, we turn to justify that the newly added weight calibration loss~$\mathcal{L}_{\rm cal}$ does not conflict with the original recommendation loss $\mathcal{L}_{\rm rec}$. 
In particular, two losses are non-conflict if the inner product of their gradients is non-negative~\cite{YuK0LHF20}. 
Technically, to justify the model convergency, we can calculate the gradients of $\mathcal{L}_{\rm rec}$ and $\mathcal{L}_{\rm cal}$ on the weight matrix~$\boldsymbol{W}$ and prove that their inner product is non-negative. The detailed proofs are as follows.

\begin{itemize}[leftmargin=5mm]
\item \textbf{Argument 1.} $\frac{\partial\mathcal{L}_{\rm rec}}{\partial \boldsymbol{W}}  \cdot \frac{\partial\mathcal{L}_{\rm cal}}{\partial \boldsymbol{W}} \geq 0.$
\item \textbf{Proof 1.} The partial derivative of the recommendation loss $\mathcal{L}_{\rm rec}$ with respect to the weight matrix~$\boldsymbol{W}$ is,
\begin{subequations}
\label{eqn_L_rec_dev}
\begin{align}
&\frac{\partial\mathcal{L}_{\rm rec}}{\partial \boldsymbol{W}} = \sum_{i=1}^{|\mathcal{I}|} \frac{\partial\mathcal{L}_{\rm rec}}{\partial \boldsymbol{w}_i} \nonumber\\
=& \sum_{i=1}^{|\mathcal{I}|} \frac{\partial\Big(-{\rm log}\big(\sigma(y_{ui} - y_{uk})\big)\Big)}{\partial \boldsymbol{w}_i} \nonumber \\
=&\sum_{i=1}^{|\mathcal{I}|} \frac{\partial\Big(-{\rm log}\big(\sigma(\boldsymbol{w}_i (\boldsymbol{e}_u \odot \boldsymbol{e}_i) - \boldsymbol{w}_k (\boldsymbol{e}_u \odot \boldsymbol{e}_k))\big)\Big)}{\partial \boldsymbol{w}_i}
\nonumber \\
=&\sum_{i=1}^{|\mathcal{I}|} \frac{\partial\Big(-{\rm log}\big(\sigma(\boldsymbol{w}_i A + B)\big)\Big)}{\partial \boldsymbol{w}_i}
\\
=&\sum_{i=1}^{|\mathcal{I}|} - A \big(1-\sigma(\boldsymbol{w}_i A + B)\big) \\
\leq& 0,
\end{align}
\end{subequations}
where we replace variables that are unrelated to the targeted parameter~$\boldsymbol{w}_i$ with constants in Eqn.~(11a), i.e, $A=\boldsymbol{e}_u \odot \boldsymbol{e}_i$ and $B=- \boldsymbol{w}_k (\boldsymbol{e}_u \odot \boldsymbol{e}_k)$. Eqn.~(11b) is derived according to the derivative of the sigmoid function, i.e., $\frac{\partial \sigma(x)}{\partial x} = \sigma(x)\big(1-\sigma(x)\big)$. Eqn.~(11c) is satisfied as $A$ refers to the summation of all modality-specific user ratings, which is almost non-negative, and $\big(1-\sigma(\boldsymbol{w}_i A + B)\big)>0$.

$\quad$ The partial derivative of the weight calibration loss $\mathcal{L}_{\rm cal}$ with respect to the weight matrix~$\boldsymbol{W}$ is,
\begin{subequations}
\label{eqn_L_cal_dev}
\begin{align}
&\frac{\partial\mathcal{L}_{\rm cal}}{\partial \boldsymbol{W}} = \sum_{i=1}^{|\mathcal{I}|} \frac{\partial\mathcal{L}_{\rm cal}}{\partial \boldsymbol{w}_i} \nonumber \\
=& \sum_{i=1}^{|\mathcal{I}|} \frac{\partial \Big({\rm nograd}(\gamma_{uik}) {\rm KL}({\rm nograd}(\boldsymbol{z}_{uik})||\boldsymbol{w}_i \oplus \boldsymbol{w}_k)\Big)}{\partial \boldsymbol{w}_i} \nonumber \\
=&\sum_{i=1}^{|\mathcal{I}|} \frac{\partial \Big(C \cdot {\rm KL}(D||\boldsymbol{w}_i \oplus \boldsymbol{w}_k) \Big)}{\partial \boldsymbol{w}_i}\\
=&\sum_{i=1}^{|\mathcal{I}|} \frac{\partial \Big( C \big(D\log D - D\log (\boldsymbol{w}_i \oplus \boldsymbol{w}_k)\big)\Big)}{\partial \boldsymbol{w}_i}\\
=&\sum_{i=1}^{|\mathcal{I}|} - \frac{CD}{\boldsymbol{w}_i} \leq 0, 
\end{align}
\end{subequations}
where variables ${\rm nograd}(\gamma_{uik})$ and ${\rm nograd}(\boldsymbol{z}_{uik})$ that have no gradient can be regarded as constants to the targeted parameter~$\boldsymbol{w}_i$, which are replaced with $C$ and $D$ in Eqn.~(12a), respectively. Eqn.~(12b) is derived according to the equation of the Kullback-Leibler divergence, i.e., ${\rm KL}(p||q) = p\log p - p\log q$. Eqn.~(12c) is satisfied since that $C, D$ and $\boldsymbol{w}_i$ are non-negative. 

$\quad$ Thereafter, the inner product between gradients of losses $\mathcal{L}_{\rm rec}$ and $\mathcal{L}_{\rm cal}$ can be calculated as follows, 
\begin{equation}
\nonumber
\label{eqn_dev_mm}
\begin{aligned}
&\frac{\partial\mathcal{L}_{\rm rec}}{\partial \boldsymbol{W}}  \cdot \frac{\partial\mathcal{L}_{\rm cal}}{\partial \boldsymbol{W}} \\
=& \Big(\sum_{i=1}^{|\mathcal{I}|} - A \big(1-\sigma(\boldsymbol{w}_i A + B)\big) \Big) \cdot \Big(\sum_{i=1}^{|\mathcal{I}|} - \frac{CD}{\boldsymbol{w}_i} \Big)  \\
\geq& 0. \qquad \qquad\qquad\qquad\qquad\qquad\quad\qquad\qquad\qquad\square
\end{aligned}
\end{equation}

\end{itemize}

\section{Experiments}
We first give the experimental settings in Subsection~\ref{section_settings}, and then show the experiment results to answer the following research questions:
\begin{itemize}[leftmargin=5mm]
\item RQ1. How effective is the proposed MARGO?
\item RQ2. What is the contribution of each component?
\item RQ3. How does MARGO perform with different hyper-parameter settings?
\item RQ4. How is the quality of learned modality weights?
\item RQ5. What is the difference between reliable and unreliable modality data?
\end{itemize}

\subsection{Experimental Settings}
\label{section_settings}
In this subsection, we introduce the detailed experimental settings, including the datasets, evaluation protocols, and implementation details.

\textbf{Datasets.} 
Following the studies~\cite{mgcn, mmsurvey}, we conduct experiments on three categories in the widely-used Amazon dataset~\cite{amazondataset}: Baby, Sports, and Clothing. To keep the fair comparison, we closely follow the pre-processing of the datasets in the study~\cite{mmsurvey}, where only the users and items that have more than 5 interactions are remained. The statistics of the three datasets after pre-processing are listed in Table~\ref{table_dataset}. In the three datasets, each item is associated with its image and text. 
We directly adopt the released modality features in the studies~\cite{mgcn, mmsurvey}, where the image feature is a $4096$-dimensional vector extracted by a CNN model~\cite{lattice} and the text feature is a $384$-dimensional vector extracted by a pre-trained sentence-transformers~\cite{ReimersG19}.

\textbf{Evaluation Metrics.}  
Following studies~\cite{mgcn, mmsurvey}, we randomly split each dataset with a ratio of $8:1:1$ to derive the training, validation, and testing sets and evaluate our proposed MARGO method in the top-$K$ recommendation task. 
To be more specific, we rank the positive items from all items in the dataset and adopt the widely-used evaluation metrics: Recall@$K$ and NDCG@$K$, to evaluate the effectiveness of our proposed method. By default, we set $K=10, 20$. 

\textbf{Implementation Details.} 
We set the embedding size~$d$ to $64$ following most studies~\cite{slmrec, DuWF0022, mgcn}. 
We tune the trade-off parameters~$\alpha$ and $\beta$ from $[0.01,0.1,1]$, respectively, and the scaling parameter~$\tau$ from $[0.1, 1, 5, 10]$. We adopt the state-of-the-art architecture in DRAGON~\cite{dragon} as the multimodal recommendation model~$\mathcal{F}$ in Eqn.~(\ref{eqn_embedding}). 
We adopt the Adam optimizer~\cite{adamoptimizer} and use the learning rate of $1e^{-4}$. The mini-batch sizes are $2,048$ for all the datasets, and the model is trained with $100$ epochs using the early stop strategy. Note that $100$ epochs are enough for the model to converge.
As same as study~\cite{mgcn, mmsurvey}, the reported results are selected according to Recall@$20$ on the validation set.

\begin{table}[t]\small
\centering
\renewcommand{\arraystretch}{1.}
\setlength{\abovecaptionskip}{0.cm}
\caption{Statistics of the experimental datasets.} 
\label{table_dataset}
\setlength{\tabcolsep}{2.mm}{
\begin{tabular}{crrcr}	
\hline
Dataset& \#User & \#Item & \#Interaction & Sparsity \\\hline
Baby & 19,445 & 7,050 & 160,792 & 99.88\% \\
Sports & 35,598 & 18,357 & 296,337 & 99.95\% \\
Clothing & 39,387 & 23,033 & 278,677 & 99.97\% \\\hline
\end{tabular}}	
\end{table}

\begin{table*}[!t]\small
\setlength{\abovecaptionskip}{0.cm}
\caption{Results of the performance comparison on the three datasets. We highlight the best and second-best results in bold and underlined, respectively. 
``impro.'' is the relative improvement. The results of the baselines are derived from their papers.}
\center
\renewcommand{\arraystretch}{1.}
\setlength{\tabcolsep}{1.3mm}{
\begin{tabular}{c|cccc|cccc|cccc|c}
\hline
& \multicolumn{4}{c|}{Baby} & \multicolumn{4}{c|}{Sports} & \multicolumn{4}{c|}{Clothing} & \multirow{2}{*}{Average}\\ \cline{2-13} 
& R@10 & R@20 & N@10 & N@20 & R@10 & R@20 & N@10 & N@20 & R@10 & R@20 & N@10 & N@20 & \\ \hline
VBPR (AAAI'16) & 0.0423 & 0.0663 & 0.0223 & 0.0284 & 0.0558 & 0.0856 & 0.0307 & 0.0384 & 0.0280 & 0.0414 & 0.0159 & 0.0193 & 0.0395\\
GRCN (MM'20) & 0.0532 & 0.0824 & 0.0282 & 0.0358 & 0.0559 & 0.0877 & 0.0306 & 0.0389 & 0.0424 & 0.0650 & 0.0225 & 0.0283 & 0.0476 \\
SLMRec (TMM'22) & 0.0540 & 0.0810 & 0.0285 & 0.0357 & 0.0676 & 0.1017 & 0.0374 & 0.0462 & 0.0452 & 0.0675 & 0.0247 & 0.0303 & 0.0517\\
MICRO (TKDE'22) & 0.0584 & 0.0929 & 0.0318 & 0.0407 & 0.0679 & 0.1050 & 0.0367 & 0.0463 & 0.0521 & 0.0772 & 0.0283 & 0.0347 & 0.0560 \\
BM3 (WWW'23) & 0.0564 & 0.0883 & 0.0301 & 0.0383 & 0.0656 & 0.0980 & 0.0355 & 0.0438 & 0.0421 & 0.0625 & 0.0228 & 0.0280 & 0.0510\\
MGCN (MM'23) & 0.0620 & 0.0964 & 0.0339 & 0.0427 & 0.0729 & 0.1106 & 0.0397 & 0.0496 & 0.0641 & 0.0945 & 0.0347 & 0.0428 & 0.0620 	\\
FREEDOM (MM'23) &  0.0627 & 0.0992 & 0.0330 & 0.0424 & 0.0717 & 0.1089 & 0.0385 & 0.0481 & 0.0629 & 0.0941 & 0.0341 & 0.0420 & 0.0615 	\\
LGMRec (AAAI'24) & 0.0644 & 0.1002 & \underline{0.0349} & \underline{0.0440} & 0.0720 & 0.1068 & 0.0390 & 0.0480 & 0.0555 & 0.0828 &0.0302 & 0.0371 & 0.0596\\ 
DRAGON (ECAI'23) & \underline{0.0662} & \underline{0.1021} & 0.0345 & 0.0435 & \underline{0.0749} & \underline{0.1124} & \underline{0.0403} & \underline{0.0500} & \underline{0.0650} & \underline{0.0957} & \underline{0.0357} & \underline{0.0435} & \underline{0.0637}\\ 
MARGO (ours) & \textbf{0.0677} & \textbf{0.1032} & \textbf{0.0362} & \textbf{0.0452} & \textbf{0.0787} & \textbf{0.1175} & \textbf{0.0430} & \textbf{0.0530} & \textbf{0.0667} & \textbf{0.0969} & \textbf{0.0364} & \textbf{0.0440} & \textbf{0.0657}\\ \hline
impro. & 2.27\% & 1.08\% & 3.72\% & 2.73\% & 5.07\% & 4.54\% & 6.70\% & 6.00\% & 2.62\% & 1.25\% & 1.96\% & 1.15\% & 3.26\% \\ \hline
\end{tabular}
}
\label{table_comparison}
\end{table*}

\subsection{Performance Comparison (RQ1)}
We select the following state-of-the-art methods to evaluate the effectiveness of the proposed method,
\begin{itemize}[leftmargin=4mm]
\item \textbf{VBPR}~\cite{vbpr} uses the item images to enrich the item embeddings and adopts the BPR loss for recommendation. For the fair comparison, we further incorporate the item texts.
\item \textbf{GRCN}~\cite{grcn} constructs a user-item interaction graph for each modality and cuts off the false-positive edges in the graph to refine the graph learning. 
\item \textbf{SLMRec}~\cite{slmrec} utilizes the self-supervised learning techniques in the graph-based models to uncover the hidden signals from the data with contrastive loss.
\item \textbf{BM3}~\cite{ZhouZLZMWYJ23} simplifies the self-supervised multimodal recommendation framework by removing the randomly sampled negative examples, while it directly perturbs the representation through a dropout mechanism.
\item \textbf{MICRO}~\cite{ZhangZLZWW23} utilizes the user-item interaction graph to learn the user and item embeddings. It then learns the item-item graph from item multimodal data, and updates the item embeddings on the item-item graph.
\item \textbf{MGCN}~\cite{mgcn} uses the item behavior information to purify the item modality features. It designs a behavior-aware fusion method to filter out the redundant information between modality features.
\item \textbf{FREEDOM}~\cite{freedom} introduces a new item-item graph for each modality and gets the latent item graphs by aggregating all the modalities. It also introduces the degree-sensitive edge pruning to denoise the user-item interaction graph.
\item \textbf{DRAGON}~\cite{dragon} conducts graph learning on both heterogeneous graph and homogeneous graphs to obtain the user and item embeddings. Besides, it employs the attentive concatenation to fuse the multimodal ratings. 
\item \textbf{LGMRec}~\cite{GuoL0WSR24} introduces local graphs to learn collaborative-related and modality-related embeddings. It further designs a global hyper-graph embedding module to capture the user's global interests.
\end{itemize}

The results of the baselines and the proposed MARGO on three datasets are shown in Table~\ref{table_comparison}, where the best and second-best results are highlighted in bold and underlined, respectively. 
We also calculate the relative improvement of MARGO compared with the best baseline in the ``impro.'' row of Table~\ref{table_comparison}.
As can be seen, the proposed MARGO method achieves the best performance compared with all baselines across all metrics. The average improvement is $3.26\%$, which demonstrates the effectiveness of the proposed method. This is because that we use the modality weights to decrease the negative effects of  unreliable modalities in the multimodal fusion. 
Besides, although several studies, e.g., DRAGON, also adds the modality weights during the multimodal recommendation, they perform worse than the proposed MARGO. This is because that MARGO involves explicit supervisions on the modality weight learning, which makes the learned weights more precise. 

\subsection{Ablation Study (RQ2)}
\label{section_ablation}
To investigate the effects of various factors, we perform ablation studies on both key components of the proposed MARGO and different modality features.

\subsubsection{Effect of Components}
To evaluate the effects of the key components, we set up the following model variants:
\begin{itemize}[leftmargin=4mm]
\item  \textit{w/o weight}. This variant removes the modality weights, where the final user rating is simply calculated by the summation of all ratings regarding different modalities and the model is trained with only the recommendation loss.
\item \textit{w/o $\mathcal{L}_{\rm cal}$}. This variant discards the proposed weight calibration loss~$\mathcal{L}_{\rm cal}$ where the modality weights are updated with only the recommendation loss. 
\item \textit{w/o two-stage}. This variant only utilizes the second training stage of MARGO to optimize the model, where the first pre-training stage is discarded. In particular, in this variant, we adopt the same framework of the proposed method, but directly optimize the model by~$\mathcal{L}_{\rm II}$ in Eqn.~(\ref{eqn_final_loss_2}).
\item \textit{w/o nograd}. This variant removes the stop-gradient operator in the weight calibration loss $\mathcal{L}_{\rm cal}$ in Eqn.~(\ref{eqn_KL_loss_refine}).
\end{itemize}

The results of the ablation study on the components are shown in Table~\ref{table_ablation}, where the best performance is highlighted in bold. We have the following observations. 

\begin{enumerate}[leftmargin=5mm]  
\item[1)] \textit{w/o weight} achieves the worst performance compared with others that involve the modality weights. This demonstrates that in multimodal recommendation, it is necessary to consider different contributions of different modalities on predicting the final user rating. 
\item[2)]  \textit{w/o $\mathcal{L}_{\rm cal}$} adds the modality weights but still achieves unsatisfied performance. This is because that in \textit{w/o $\mathcal{L}_{\rm cal}$}, the modality weights only updated with the recommendation loss, without explicit supervisions. Therefore, the learned modality weights may be imprecise.
\item[3)] \textit{w/o two-stage} achieves worse performance than MARGO that adopts the two-stage training process. This may be because that without the first stage, the predicted user ratings at initial training epochs are irrational. Therefore, the derived modality reliability vector cannot provide correct supervisions to the weight learning, and hence harm the fusion results. This demonstrates the necessity of the two-stage training process.
\item[4)]  \textit{w/o nograd} results in a little performance drop compared with our proposed MARGO. Although the differences are not that obvious, we empirically find that \textit{w/o nograd} will hurt the diversity of the learned modality weights, which will be discussed in RQ4 (Subsection \ref{section_weight_visualization}).
\end{enumerate}

\begin{table*}[!th]\small
\setlength{\abovecaptionskip}{0.0cm}
\caption{Results of the ablation study on both the modules of the proposed method and modality features.}
\center
\renewcommand{\arraystretch}{1.}
\setlength{\tabcolsep}{2.2mm}{
\begin{tabular}{c|cccc|cccc|cccc}
\hline
 & \multicolumn{4}{c|}{Baby} & \multicolumn{4}{c|}{Sports} & \multicolumn{4}{c}{Clothing} \\ \cline{2-13} 
 & R@10 & R@20 & N@10 & N@20 & R@10 & R@20 & N@10 & N@20 & R@10 & R@20 & N@10 & N@20 \\ \hline
 \textit{w/o weight} & 0.0612 & 0.0935 & 0.0327 & 0.0410 & 0.0704 & 0.1045 & 0.0384 & 0.0472 & 0.0616 & 0.0908 & 0.0337 & 0.0411 \\
\textit{w/o $\mathcal{L}_{\rm cal}$} & 0.0654 & 0.1014 & 0.0357 & 0.0450 & 0.0760 & 0.1143 & 0.0413 & 0.0512 & 0.0652 & 0.0957 & 0.0360 & 0.0437 \\
\textit{w/o two-stage} & 0.0659 & 0.1018 & 0.0354  & 0.0446 & 0.0762 & 0.1145 & 0.0419 & 0.0517 & 0.0655 & 0.0948  & 0.0358 & 0.0432 \\
\textit{w/o nograd} & 0.0656 & 0.1023 & 0.0351  & 0.0446 & 0.0785 & 0.1152 & 0.0430 & 0.0525 & 0.0663 & \textbf{0.0976}  & 0.0361 & 0.0440 \\\hline
\textit{MARGO$_v$} & 0.0497 & 0.0798 & 0.0272 & 0.0350 & 0.0614 & 0.0910 & 0.0336 & 0.0413 & 0.0459 & 0.0669 & 0.0244 & 0.0297 \\
\textit{MARGO$_t$} & 0.0635 & 0.0985 & 0.0338 & 0.0428 & 0.0727 & 0.1089 & 0.0394 & 0.0488 & 0.0599 & 0.0895 & 0.0329 & 0.0405 \\ \hline
MARGO & \textbf{0.0677} & \textbf{0.1032} & \textbf{0.0362} & \textbf{0.0452} & \textbf{0.0787} & \textbf{0.1175} & \textbf{0.0430} & \textbf{0.0530} & \textbf{0.0667} & 0.0969 & \textbf{0.0364} & \textbf{0.0440} \\ \hline
\end{tabular}
}
\label{table_ablation}
\end{table*}

\subsubsection{Effect of Modalities}
\label{section_modality_test}
To explore the contribution of each modality to the recommendation performance, we compare the performance of the propose method under the uni-modal setting. \textit{MARGO$_v$} and \textit{MARGO$_t$} denote the models that utilize visual and textual features, respectively. The results of the ablation study on the modalities are shown in Table~\ref{table_ablation}. Comparing the results, we have the following observations.
\begin{enumerate}[leftmargin=5mm]  
\item[1)] The proposed MARGO that utilizes the multimodal data of items outperforms the other two variants that only utilize uni-modal data. This proves that the proposed MARGO could effectively use the item multimodal data to help improve the recommendation performance. 
\item[2)] The variant \textit{MARGO$_t$} that utilizes the textual information achieves the better performance than \textit{MARGO$_v$} that uses the visual information. This  indicates that it can better learn the user interests that using texts than images.
\item[3)] Comparing the results of the \textit{w/o weight} variant, we find that although it uses both item texts and images, it still achieves the worse performance than \textit{MARGO$_t$} that only uses the item texts in Baby and Sports datasets. This indicates that the simple multimodal fusion method will lead to the performance degradation issue. Differently, the performance of MARGO is better than both \textit{MARGO$_v$} and \textit{MARGO$_t$}, which proves that MARGO is able to settle the performance degradation issue. 
\end{enumerate}

\begin{figure*}[!th]
\centering
\setlength{\abovecaptionskip}{0.3cm}
\includegraphics[width=\linewidth]{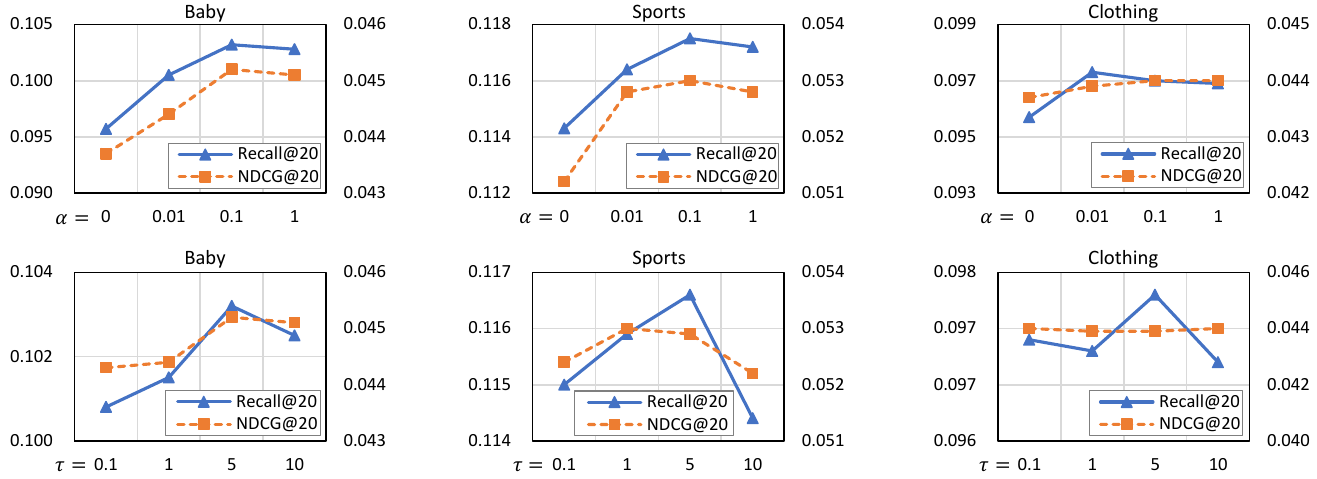}
\caption{Performance of the proposed MARGO with respect to different trade-off parameter~$\alpha$ on three datasets. The x-axis refers to the values of $\alpha$. The left and right y-axis refer to the Recall@20 and NDCG@20, respectively.}
\label{fig_para}
\end{figure*} 

\subsection{Hyper-parameter Discussion (RQ3)}
In this subsection, we evaluate the key hyper-parameters in MARGO, i.e., the trade-off parameter~$\alpha$ defined in Eqn.~(\ref{eqn_final_loss_2}) and sensitivity parameter~$\tau$ defined in Eqn.~(\ref{eqn_confidence}). 

\subsubsection{Trade-off parameter $\alpha$}
It adjusts the balance between the original recommendation loss and weight calibration loss, where a larger~$\alpha$ indicates the greater importance of the weight calibration loss in optimization. We tune $\alpha$ from $\{0,0.01,0.1,1\}$ and report the Recall@$20$ and NDCG@$20$ results of MARGO in Figure~\ref{fig_para}. 
As can be seen, when we discard the weight calibration process, i.e., $\alpha=0$, the model achieves the worst performance. This demonstrates the effectiveness of our proposed the weight calibration loss. Besides, along with $\alpha$ increasing, the model performance first rises significantly until it achieves the best performance with the most suitable $\alpha$ and then decreases. This demonstrates the necessity of appropriately adjust the importance of the weight calibration to supervise the modality weight learning in the multimodal recommendation. 
However, when~$\alpha$ becomes excessively large, the model focuses excessively on the weight learning, while overlooking the original and dominated recommendation objective, and hence the performance drops. Therefore, finding a suitable trade-off is crucial for ensuring the effectiveness of the recommendation. 

\subsubsection{Sensitivity parameter~$\tau$}
It adjusts the sensitivity of the confidence $\gamma_{uik}$ affected by the prediction results $y_{ui}-y_{uk}$ during the weight calibration process. When $\tau$ becomes large, the confidence changes smoothly with the prediction results. We tune $\tau$ from $\{0.1,1,5,10\}$ and report the Recall@$20$ and NDCG@$20$ results of MARGO in Figure~\ref{fig_para}. As can be seen, along with the sensitivity parameter~$\tau$ increasing, the performance first rises significantly until $\tau=5$ and then decreases in all datasets. This suggests that a suitable sensitivity parameter is crucial for the weight calibration.

\begin{figure*}[!t]
\centering
\setlength{\abovecaptionskip}{0.3cm}
\includegraphics[width=\linewidth]{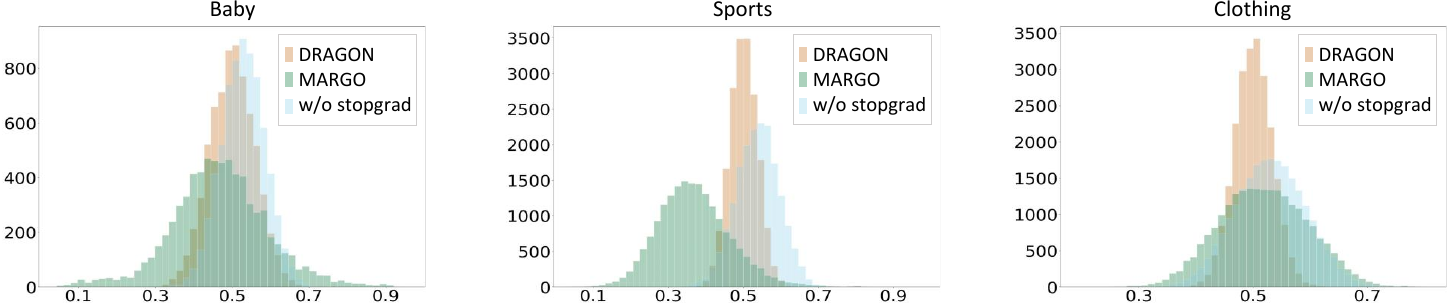}
\caption{Histogram of the weights of images learned by the best baseline DRAGON, variant \textit{w/o stopgrad}, and our proposed method MARGO. The x-axis and y-axis refer to the weight value and item number, respectively.}
\label{fig_weight}
\end{figure*}

\begin{figure*}[!th]
\centering
\setlength{\abovecaptionskip}{0.3cm}
\includegraphics[width=\linewidth]{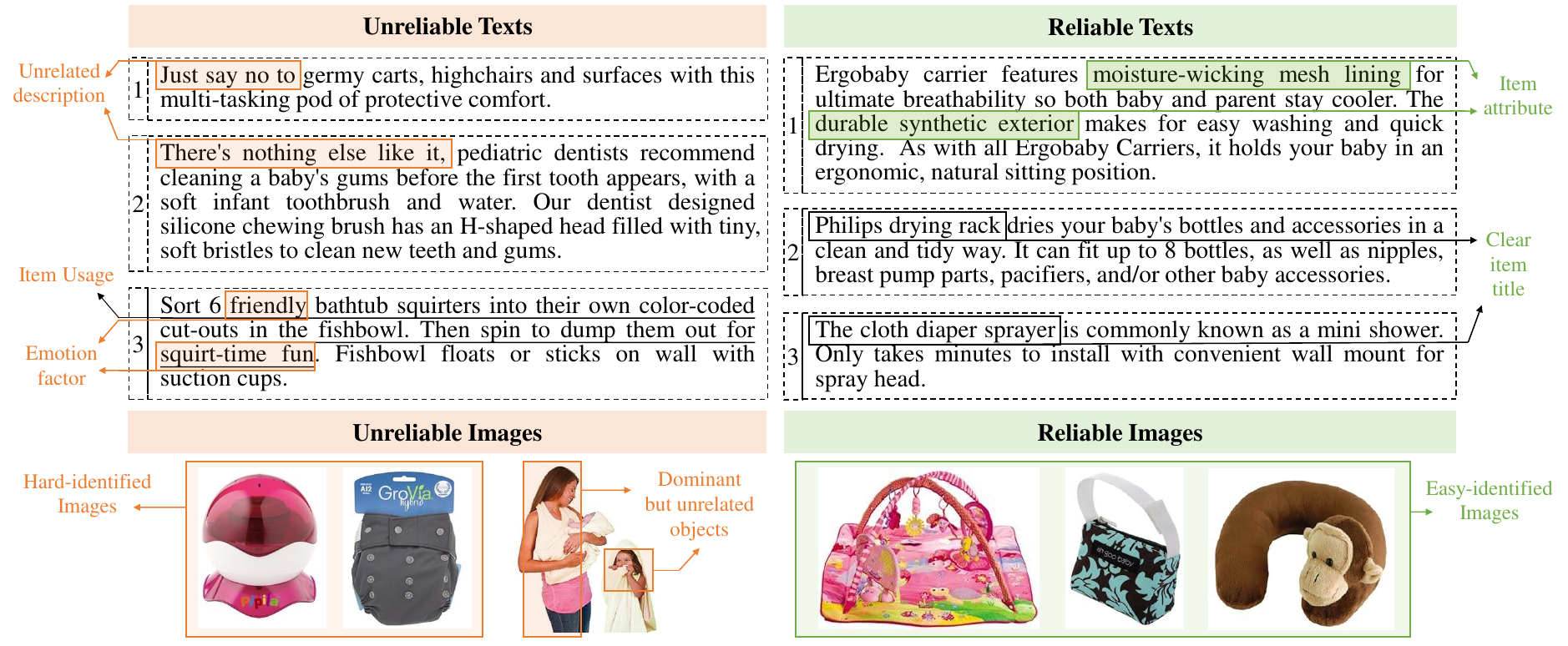}
\caption{Examples of the five most reliable and unreliable item texts and images in the Baby dataset. The reliability of the modality data is determined by the learned modality weights learned by the proposed MARGO.}
\label{fig_visualization}
\end{figure*} 

\subsection{Visualization of the Modality Weights (RQ4)}
\label{section_weight_visualization}

We visualize the weights of images learned by the best baseline (i.e., DRAGON) and our proposed MARGO in Figure~\ref{fig_weight}. Besides, we also visualize the image weights learned by the \textit{w/o nograd} variant to show it differences of the learned weights compared to our proposed MARGO. In Figure~\ref{fig_weight}, the x-axis refers to the interval of weight values, while the y-axis refers to the number of items whose weights of the image fall within the corresponding weight interval. The histogram shows the distribution of the learned weights of images. Note that the distribution of the learned weights of texts is symmetrical with that of images, since that the summation of the weights of the text and image in one item is equal to $1$. 
From Figure~\ref{fig_weight}, we have the following observations.
\begin{enumerate}[leftmargin=5mm]
\item[1)] The expectation of the weights learned by DRAGON is close to $0.5$. This may be because that DRAGON does not have specific supervisions to distinguish whether the image is reliable, whereby the model provides a neutral prediction. Differently, the expectation of the weights learned by MARGO is smaller than $0.5$. It shows that images are less reliable than texts as a whole. This phenomenon is in line with the result in Effect of Modalities (Subsection \ref{section_modality_test}), where \textit{MARGO$_t$} that utilizes the item texts outperforms \textit{MARGO$_v$} that uses the item images. 
\item[2)] The weights learned by MARGO are more dispersed than those learned by DRAGON. Specifically, the weights of images learn by DRAGON are mainly concentrated in $[0.4,0.6]$, indicating that there is small difference in weights between images and texts. The result shows that the weights learned by DRAGON fails to meet the motivation of learning different weights for modalities. Differently, the weights learned by MARGO are more dispersed than DRAGON, which demonstrates that MARGO can better capture different contributions of modalities and hence achieve better performance. 
\item[3)] Although \textit{w/o nograd} adds the supervisions on the modality weight learning as same as MARGO, the learned weights are more concentrated than MARGO. This may be because that the modality weights learned without explicit supervisions tend to be learned concentrated, according to the results of DRAGON. Therefore, without the stop-gradient operator in \textit{w/o nograd} variant, the target modality reliability vector will be affected by the modality weights, causing the target to become concentrated and lose its original diversity.
This further demonstrates that it is necessary to add the stop-gradient operator on the target modality reliability vector to avoid the negative gradients.
\end{enumerate}

\subsection{Case Study (RQ5)}
In this subsection, we conduct a case study to intuitively explore the similar and different patterns in the most reliable and unreliable modality data. In particular, without losing generality, we select most $3$ reliable and unreliable modality data in Baby dataset based on their modality weights learned by our proposed MARGO in Figure~\ref{fig_visualization}. 
From Figure~\ref{fig_visualization}, we have the following observations. 

\begin{enumerate}[leftmargin=5mm]
\item[1)] The reliable and unreliable texts are different in contents. To be more specific, unreliable texts tend to have more unrelated descriptions or spend lots of space on describing the item usage. Reliable texts often begin with the item titles and utilize more concise phrase to introduce the item functions and attributes. 
For example, in unreliable texts, the phrases ``just say no to'' and `` there's nothing else like it'' are not necessary to describe the item. The third example details how to use an item that appears to be a children's bath toy. Differently, it is clear from the first example in reliable texts that the item is a ``Ergobaby Carrier'', which has ``moisture-wicking mesh lining'' and ``durable synthetic exterior'', making it breathable, easy wash and quick dry. 
\item[2)] The reliable and unreliable texts are different in language styles, where the unreliable texts contain more emotional expression such as ``friendly'' and ``out for squirt-time fun''. These factors might mislead the model to focus on the emotion of texts, rather than the information of the item. 
\item[3)] The reliability of images mainly depends on how much could the images describe the item. To be more specific, in unreliable images, it is difficult for us to recognize what the item is used for with only the image. For example, it can be seen that the first item is a pink boll with a pedestal from the image. However, it is hard to know that it is a pacifier sterilizer until we refer to its text. Besides, as for the second image, the persons are more prominent compared to the true item, i.e., the bath towel. 
Differently, in the reliable images, the item is dominant in the image and is easier to recognize based on the images. For example, the first item is a blanket that provides the baby a place to play, and the second item is a handbag.
\end{enumerate}

\section{Conclusion and Future Work}
In this paper, we propose a modality reliability guided multimodal recommendation framework (MARGO) that leverage weights to model different contributions of item modalities in the rating prediction. Different from existing methods that optimize the weights with only the recommendation objective, we introduce a new supervision signals to enhance the weight learning process. In particular, we first derive a modality reliability vector by comparing the multimodal ratings of positive and negative items as the supervision label of the modality weights. We then calculate the confidence level for the supervision label for dynamically adjusting the supervision strength and eliminating the negative supervision. Extensive experiments on three public datasets have demonstrated the effectiveness of the proposed MARGO. 

Through the experimental analysis, we have found that the reliable and unreliable modality data shares their specific patterns. For example, reliable texts mainly describe the item attributes and functions, while unreliable texts contain more emotional adjectives.  
It is benefit for the multimodal recommendation to discover and filter out such irrelevant information  from the modality features, e.g., emotional factors in texts. Therefore, in the future, we plan to devise an automatic method for the feature enhancement for recommendation.

\ifCLASSOPTIONcompsoc
\else
\fi


\bibliographystyle{IEEEtran}
\bibliography{ref.bib}

\begin{IEEEbiography}[{\includegraphics[width=1in,height=1.25in,clip,keepaspectratio]{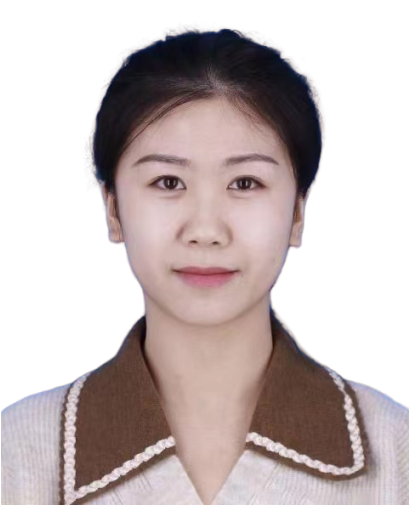}}]{Xue Dong} received the Ph.D. degree from the School of Software, Shandong University, in 2023. She is currently a research fellow with the School of Software, Tsinghua University, China. Her research interests contain multimedia computing, multimodal recommendation and multimodal prompt tuning. She has published several papers in the top venues, such as ACM SIGIR, MM, TOIS, IEEE TNNLS and Information Fusion.
\end{IEEEbiography}

\begin{IEEEbiography}[{\includegraphics[width=1in,height=1.25in,clip,keepaspectratio]{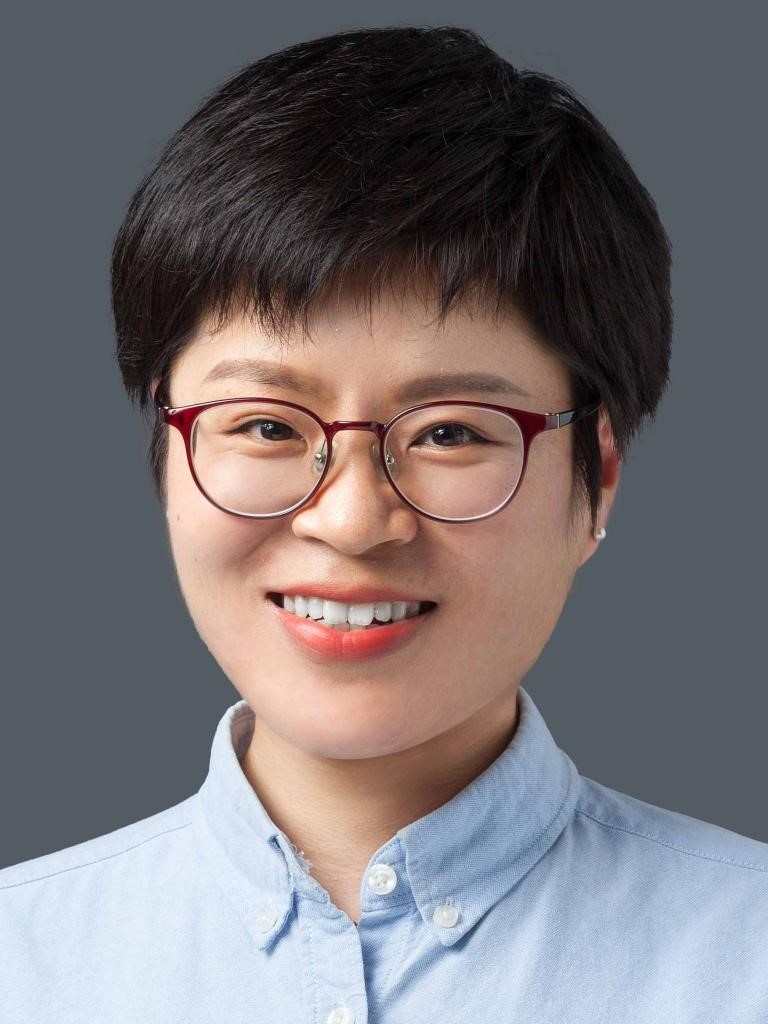}}]{Xuemeng Song} received the B.E. degree from the University of Science and Technology of China, in 2012, and the Ph.D. degree from the School of Computing, National University of Singapore, in 2016. She is currently an Associate Professor with Shandong University, China. She has published several papers in the top venues, such as ACM SIGIR, MM, and TOIS. Her research interests include information retrieval and social network analysis. She has served as a reviewer for many top conferences and journals.
\end{IEEEbiography}

\begin{IEEEbiography}[{\includegraphics[width=1in,height=1.25in,clip,keepaspectratio]{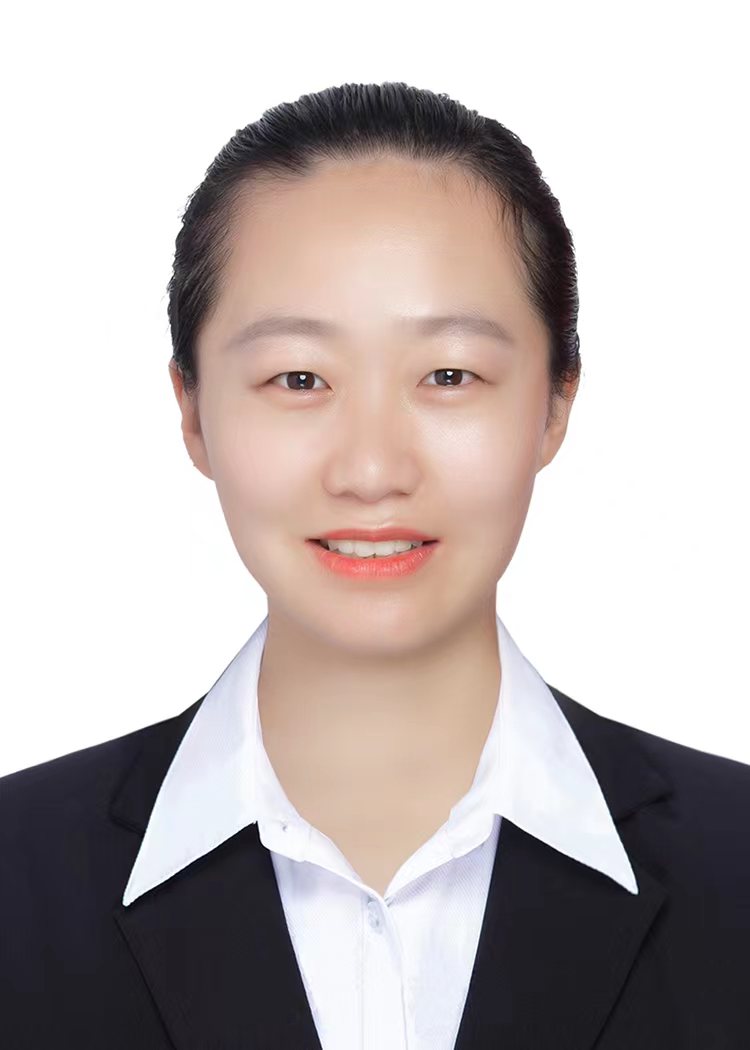}}]{Na Zheng} received the M.S. degree from the Central South University in 2018, and the Ph.D. degree from the Shandong University, in 2022. She is currently a research fellow with the National University of Singapore. She has published several papers in the top venues, such as ACM MM, TOIS, ToMM and IEEE TCVST. Her research interests include computer vision and food computing. She has served as a reviewer for many top conferences and journals.  
\end{IEEEbiography}

\begin{IEEEbiography}[{\includegraphics[width=1in,height=1.25in,clip,keepaspectratio]{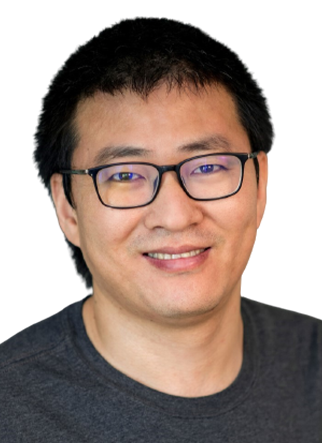}}]{Sicheng Zhao} received the Ph.D. degree from the Harbin Institute of Technology, Harbin, China, in 2016. 
He was a Visiting Scholar with the National University of Singapore, Singapore, from 2013 to 2014; a Research Fellow with
Tsinghua University, Beijing, China, from 2016 to 2017; a Postdoctoral Research Fellow with the University of California at Berkeley, Berkeley, CA, USA, from 2017 to 2020; and a Postdoctoral Research Scientist with Columbia University, New York, NY, USA, from 2020 to 2022. He is currently a Research Associate Professor with Tsinghua University. His research interests include affective computing, multimedia, and computer vision.
\end{IEEEbiography}

\begin{IEEEbiography}[{\includegraphics[width=1in,height=1.25in,clip,keepaspectratio]{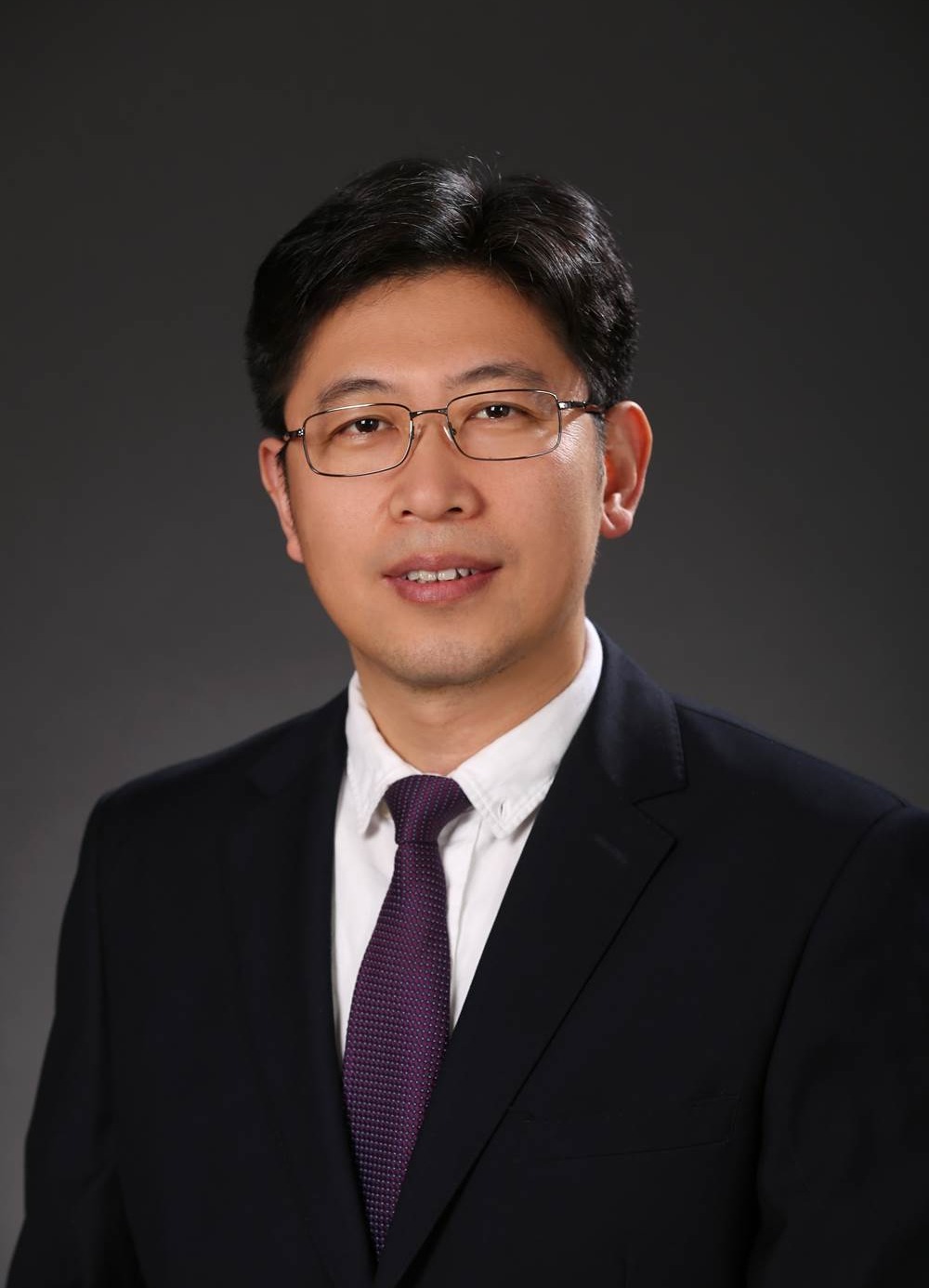}}]{Guiguang Ding} received the Ph.D. degree from Xidian University, Xi’an, China, in 2004. In 2006, he was a Postdoctoral Research Fellow with the Department of Automation, Tsinghua University, Beijing, China. He is currently a Professor with the School of Software, Tsinghua University. He has published over 100 scientific papers in major journals and conferences. His current research interests include multimedia information retrieval, computer vision, and machine learning. Dr. Ding served as a Leading Guest Editor for Neural Processing Letters (NPL) and Multimedia Tools and Applications (MTAP), the Special Session Chair for IEEE International Conference on Acoustics, Speech, and Signal Processing (ICASSP) 2021, IEEE International Conference on Multimedia and Expo (ICME) 2019 and 2020, and Pacific Rim Conference on Multimedia (PCM) 2017, and a reviewer for over 20 prestigious international journals and conferences.
\end{IEEEbiography}

\end{document}